# Free convection over a non-isothermal axisymmetric body immersed in a porous medium saturated with an electrically conducting non-Newtonian fluid

Shobha Bagai, Chandrashekhar Nishad

Department of Mathematics

University of Delhi

New Delhi, India

**Abstract**

The study investigates the problem of magnetohydrodynamic (MHD) free convection over a non-isothermal axisymmetric body under the action of transverse magnetic field. The body is embedded in a porous medium saturated with electrically conducting non-Newtonian power law fluid. In order to obtain similarity solution, it is assumed that the viscosity of the fluid decays exponentially with temperature. The qualitative results are illustrated for a vertical flat plate, horizontal cylinder and sphere.

1. **Introduction**

Magnetohydrodynamic (MHD) free convection flow of electrically conducting fluids through different geometries is of considerable interest to the technical field due to its frequent occurrence in geothermal operations, petroleum industries, thermal insulation and design of solid-matrix heat exchangers chemical catalytic reactor. As an example, the geothermal region gases are electrically conducting and undergo the influence of magnetic field. It also has applications in nuclear engineering in connection with reactors cooling. Cowling [1] provides an extensive study of application of magnetohydrodynamics to geophysical and astronomical problems. References of comprehensive surveys regarding the subject of porous media can be had in the recent books [2 – 5].

Most of the studies concerned with the problem of MHD free convection flow in non-porous and porous media have been published during the past several decades. Although, MHD free convection flow of an electrically conducting fluid across a vertical plate in a non-porous medium have been considered by various authors [6 – 10], very few authors have considered the problem for axisymmetric bodies. The problem of MHD free convection flow of an electrically conducting fluid flow in porous medium across vertical plate without internal heat generation has been studied by Raptis and Perdikis [11]. Al-Nimr and Hader [12] have presented analytical solutions for fully developed MHD natural convection flow in open ended vertical porous channels. An analysis for non-Darcy free convection flow of an electrically conducting fluid over an impermeable vertical plate embedded in a thermally stratified, fluid saturated porous medium for the case of power-law surface temperature was presented by Afify [13]. A handful of works are available in the literature for MHD free convection flow in porous medium for different geometries with internal heat generation. Yih [14] numerically analyses the effects of viscous dissipation, Joule heating and heat source/ sink on non-Darcy MHD natural convection flow over an isoflux permeable sphere in a porous medium. The MHD free convection from a sphere embedded in an electrically conducting fluid saturated porous regime with heat generation was examined theoretically and numerically by Beg et al [15].

In the present paper we study the effect of variable temperature dependent viscosity on MHD free convective heat transfer rates in the presence of internal heat generation for a non-isothermal two dimensional or axisymmetric bodies embedded in porous medium saturated with electrically conducting non-Newtonian fluid. Similarity solutions are obtained for exponentially decaying viscosity of the fluid. For the existence of the similarity solution, the wall temperature distribution is taken as an arbitrary power of the wall profile parameter $\xi$. Although the equations are derived for an arbitrary axisymmetric shape, numerical results are illustrated only for the three geometries namely, vertical flat plate, horizontal cylinder and sphere.

2. **Mathematical Formulation**

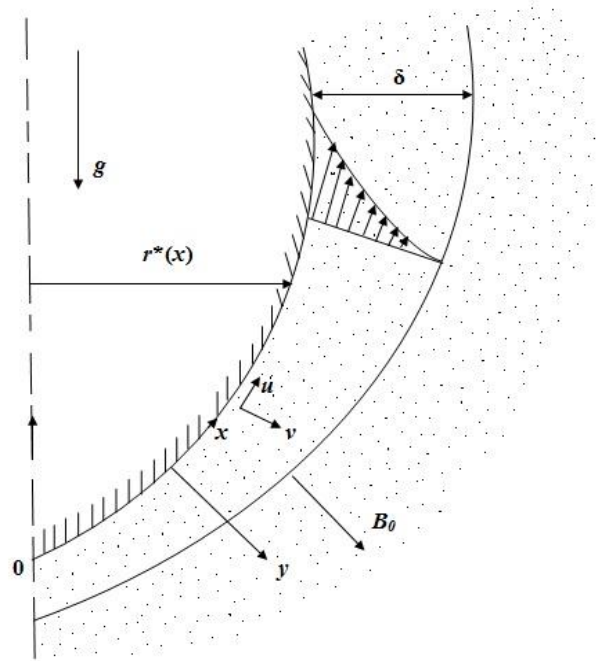

Figure 1 Physical model

Consider the problem of free convection boundary- layer flow over a non-isothermal axisymmetric body embedded in a non-Newtonian power law fluid saturated porous medium as shown in Figure 1. A strong magnetic field of uniform strength $B_0$ is applied in the $y$- direction normal to the body surface. The fluid is assumed to be electrically conducting. The induced magnetic field due to motion of electrically conducting fluid is neglecting. This assumption is valid because the velocity of the fluid is very small in porous medium. The body is maintained at a wall temperature $T_w(x)$ which is higher than the ambient temperature $T_\infty$ everywhere.

Under the Boussinesq approximation, the equations for MHD free convection in a curvilinear co-ordinate system can be written as [2]

$$\frac{\partial}{\partial x}(r^* u) + \frac{\partial}{\partial y}(r^* v) = 0 \qquad (1)$$

$$u^n = \frac{K^*(n) g_x \beta \Delta T_w}{\upsilon^*} - \frac{K^*(n) \sigma_0 B_0^{\,2}}{\upsilon^* \rho \varepsilon} u \qquad (2)$$

$$u \frac{\partial T}{\partial x} + v \frac{\partial T}{\partial y} = \alpha_m \frac{\partial^2 T}{\partial y^2} + \frac{q'''}{\rho c_{pf}} \qquad (3)$$

where the function $r^*(x)$ describes the geometric configuration of the body and is given as

$$r^*(x) = \begin{cases} 1 & : \text{plane flow} \qquad 4(a) \\ r(x) & : \text{axisymmetric flow} \qquad 4(b) \end{cases}$$

The acceleration due to gravity $g_x$ along the x-component is related to the body shape function $r^*(x)$ as

$$g_x = g \left[ 1 - \left( \frac{dr^*}{dx} \right)^2 \right]^{\frac{1}{2}} \qquad (5)$$

The power law fluid index $n$ is less than 1 for pseudoplastic fluids, equal to 1 for Newtonian fluids and greater than 1 for dilatant fluids. The non-Newtonian power laws proposed by Christopher and Middleman [16] and Dharmadhikari and Kale [17] have a common form namely

$$-\frac{dp}{dx} - \frac{\mu^*}{K^*(n)} u^n - \rho g_x = 0 \qquad (6)$$

where $\mu^*$ is dynamic viscosity parameter while $K^*(n)$ is the modified permeability given as a function of the power law index $n$.

The appropriate boundary conditions associated with the problem are

$$\text{At } y = 0 : v = 0, \quad T = T_w(x) \qquad (7)$$

$$\text{As } y \to \infty : u \to 0, \quad T \to T_\infty \tag{8}$$

We introduce the general transformations

$$\psi = A(x) f(\eta) \tag{9}$$

$$\eta = B(x) y \tag{10}$$

$$T - T_\infty = \Delta T_w(x) \theta(\eta) \tag{11}$$

$$\upsilon^* = \upsilon_0^* e^{-b(T-T_\infty)}, \text{ where } \upsilon_0^* \text{ is the viscosity at } T = T_\infty \tag{12}$$

where $\psi$ is the stream function, which satisfies the equation of continuity (1)

$$u = \frac{1}{r^*} \frac{\partial \psi}{\partial y}, \quad v = -\frac{1}{r^*} \frac{\partial \psi}{\partial x} \tag{13}$$

On using equations (9) – (12), equations (2) and (3) reduce to

$$(f'(\eta))^n = \left[ \left( \frac{r^* \alpha_m Ra_x}{xA(x)B(x)} \right)^n \theta(\eta) - \frac{K^*(n)\sigma_0 B_0^2}{\upsilon^* \rho \varepsilon} \left( \frac{r^*}{A(x)B(x)} \right)^{(n-1)} f'(\eta) \right] e^{\gamma \theta(\eta)} \tag{14}$$

$$P(x) f'(\eta) \theta(\eta) - M(x) f(\eta) \theta'(\eta) = (\alpha_m r^* x) \left( \frac{B(x)}{A(x)} \right) \theta''(\eta) + \frac{x^2}{\Delta T_w Ra_x \alpha_m} \frac{q'''}{\rho c_{pf}} \tag{15}$$

where

$$\Delta T_w(x) = T_w(x) - T_\infty$$

$$Ra_x = \left( \frac{K^*(n) g_x \beta \Delta T_w(x) x^n}{\alpha_m^n \upsilon_0^*} \right)^{\frac{1}{n}}$$

$$P(x) = \frac{d \ln \Delta T_w(x)}{d \ln x} \tag{16}$$

$$M(x) = \frac{d \ln A(x)}{d \ln x}$$

$\gamma = b \Delta T_w(x)$ is the dimensionless viscosity parameter.

The functions $A(x)$ and $B(x)$ are chosen so as to satisfy

$$\frac{r^{*}\alpha_{m}Ra_{x}}{xA(x)B(x)} = 1 \qquad (17)$$

$$(\alpha_{m}r^{*}x)\frac{B(x)}{A(x)} = \frac{1}{I(x)} \qquad (18)$$

Simplifying (17) and (18) yields

$$I(x) = \frac{\int\limits_{0}^{x} (\Delta T_{w}(x))^{\left(\frac{(s-1)n+1}{n}\right)} r^{*2} g_{x}^{\frac{1}{n}} dx}{(\Delta T_{w}(x))^{\left(\frac{(s-1)n+1}{n}\right)} r^{*2} g_{x}^{\frac{1}{n}} x}$$

$$A(x) = \alpha_{m}r^{*}(Ra_{x}I(x))^{\frac{1}{2}} \qquad (19)$$

$$B(x) = \frac{1}{x}\left(\frac{Ra_{x}}{I(x)}\right)^{\frac{1}{x}}$$

On using equation (19), equations (14) and (15) can be written as

$$(f'(\eta))^{n} = [\theta(\eta) - Nf'(\eta)]e^{\gamma\theta(\eta)} \qquad (20)$$

$$\theta''(\eta) + \frac{1}{2}[1-(s-1)P(x)I(x)]f(\eta)\theta'(\eta) - P(x)I(x)f'(\eta)\theta(\eta) + \frac{1}{(1+s\lambda)}e^{\frac{-\eta}{(1+s\lambda)^{\frac{1}{2}}}} = 0 \qquad (21)$$

The non-Negative nom-dimensional MHD parameter $N$ is given as

$$N = Ra_{x}\left(\frac{\alpha_{m}\sigma_{0}B_{0}^{2}}{x\rho\varepsilon g_{x}\beta\Delta T_{w}(x)}\right) \qquad (22)$$

whereas the internal heat generation per unit volume $q'''$ is written as [18]

$$q''' = \frac{k_{m}\Delta T_{w}(x)}{x^{2}(1+s\lambda)}\left(\frac{Ra_{x}}{I(x)}\right)e^{\frac{-\eta}{(1+s\lambda)^{\frac{1}{2}}}} \qquad (23)$$

The primes in equation (20) and (21) denote differentiation with respect to the similarity variable $\eta$. The appropriate boundary conditions now associated with the problem in terms of the similarity variable $\eta$ are

$$\text{At } \eta = 0 : \quad f = 0, \quad \theta = 1$$
$$\text{As } \eta \to \infty : \quad f \to 0, \quad \theta \to 0 \tag{24}$$

The transformed Darcian velocities along the body and in the normal direction are given as

$$u = \frac{\alpha_m Ra_x}{x} f'(\eta) \tag{25}$$

$$v = \left(\frac{\alpha_m}{x}\right)\left(\frac{Ra_x}{I(x)}\right)^{\frac{1}{2}} \left[\left\{(s-1)P(x)I(x) - \frac{1}{2}\right\} f(\eta) \right.$$
$$\left. -\left\{\frac{(s-1)n+2}{2n} P(x)I(x) + \frac{I(x)}{n}\frac{d \ln g_x}{d \ln x} + I(x)\frac{d \ln r^*}{d \ln x} - \frac{1}{2}\right\} \eta f'(\eta)\right] \tag{26}$$

### 3. Solution Procedure

Equations (20) and (21) suggest that exact solutions are possible when the lumped parameter $P(x)I(x)$ remains constant in the stream wise direction. The lumped parameter $P(x)I(x)$ can be written as

$$P(x)I(x) = \frac{d \ln \Delta T_w(x)}{d \ln x} \frac{\int_0^x (\Delta T_w(x))^{\left(\frac{(s-1)n+1}{n}\right)} r^{*2} g_x^{\frac{1}{n}} dx}{(\Delta T_w(x))^{\left(\frac{(s-1)n+1}{n}\right)} r^{*2} g_x^{\frac{1}{n}} x} \tag{27}$$

where $s$ is a non-negative integer.

On introducing a new transformed variable

$$\xi = \int_0^x r^{*2} g_x^{\frac{1}{n}} dx \tag{28}$$

equation (27) can be written as

$$P(x)I(x) = \frac{d \ln \Delta T_w(x)}{d \ln \xi} \frac{\int_0^\xi (\Delta T_w(x))^{\left(\frac{(s-1)n+1}{n}\right)} d\xi}{(\Delta T_w(x))^{\left(\frac{(s-1)n+1}{n}\right)} \xi} \tag{29}$$

The variable $\xi$ is the distance measured from the lower stagnation point for the case of plane flow and the volume segment cut by a horizontal plane for the case of axisymmetric flow. Equation (29) reveals that the similarity solutions exists when the wall temperature varies according to,

$$\Delta T_w(x) \propto \xi^\lambda \qquad (30)$$

Hence, the lumped parameter $P(x)I(x)$ is related to the constant exponent $\lambda$, power index $n$ and non-negative integer $s$ by the relation

$$P(x)I(x) = \frac{\lambda n}{((s-1)n+1)\lambda + n} \qquad (31)$$

Equation (21) therefore takes the form

$$\theta''(\eta) + \frac{\lambda + n}{2[((s-1)n+1)\lambda + n]} f(\eta)\theta'(\eta) - \frac{\lambda n}{((s-1)n+1)\lambda + n} f'(\eta)\theta(\eta) + \frac{1}{(1+s\lambda)} e^{\frac{-\eta}{(1+s\lambda)^{\frac{1}{2}}}} = 0 \qquad (32)$$

On using equations (19), (28) and (30), the function $I(x)$ in equation (19) can be written as

$$I(x) = \frac{n}{((s-1)n+1)\lambda + n} \frac{\xi}{g_x^{\frac{1}{n}} r^{*2} x} \qquad (33)$$

The transformed governing equations are valid for an axisymmetric body of arbitrary shape. The geometric configuration is intrinsic due to the transformed variable $\xi$. Three geometric shapes are considered namely vertical flat plate, horizontal cylinder and sphere. For these geometries the variable $\xi$, from equation (28) is given by,

$$\xi = \begin{cases} g^{\frac{1}{n}} x & \text{: vertical flat plate} \\ g^{\frac{1}{n}} r \int_0^\phi (\sin\phi)^{\frac{1}{n}} d\phi & \text{:horizontal cylinder} \\ g^{\frac{1}{n}} r^3 \int_0^\phi (\sin\phi)^{\frac{2n+1}{n}} d\phi & \text{:sphere} \end{cases} \qquad (34)$$

where $\phi = \sin^{-1}\left(\frac{x}{r}\right)$ and $r$ is the radius of the horizontal cylinder or sphere.

The local surface heat flux is given by

$$q_w = -k_m \left.\frac{\partial T}{\partial y}\right|_{y=0} \qquad (35)$$

whereas the non dimensional local heat flux defined by

$$q^* = \left(\frac{q_w L_r}{\Delta T_{wr} k_m}\right)\left[\left(\frac{L_r}{\alpha_m}\right)\left(\frac{K^*(n) g_x \beta \Delta T_{wr}}{\upsilon_0^*}\right)^{\frac{1}{n}}\right]^{-\frac{1}{2}} \tag{36}$$

$\Delta T_{wr}$ is the wall ambient temperature difference at a trailing edge or rear stagnation point and $L_r$ is the reference length such as the length of the vertical plate or radius of the cylinder or sphere.

The local Nusselt number defined by

$$Nu_x = \frac{q_w x}{\Delta T_w(x) k_m} \tag{37}$$

can be written using equation (35) as

$$Nu_x = -\theta'(0)\left(\frac{Ra_x}{I(x)}\right)^{\frac{1}{2}} \tag{38}$$

Equation (36) for three geometries under consideration can be written as

$$q^* = -\theta'(0)\left(\frac{x}{L_r}\right)^{\frac{(2n+1)\lambda - n}{2n}}\left\{\frac{((s-1)n+1)\lambda + n}{n}\right\}^{\frac{1}{2}} \quad \text{vertical flat plate} \tag{39}$$

$$q^* = -\theta'(0)(\sin\phi)^{\frac{1}{n}}\left[\frac{\left(\int_0^\phi (\sin\phi)^{\frac{1}{n}} d\phi\right)^{\frac{(2n+1)\lambda - n}{2n}}}{\left(\int_0^\pi (\sin\phi)^{\frac{1}{n}} d\phi\right)^{\frac{(2n+1)\lambda}{2n}}}\right]\left\{\frac{((s-1)n+1)\lambda + n}{n}\right\}^{\frac{1}{2}} \quad \text{horizontal cylinder} \tag{40}$$

$$q^* = -\theta'(0)(\sin\phi)^{\frac{2n+1}{2n}}\left[\frac{\left(\int_0^\phi (\sin\phi)^{\frac{2n+1}{n}} d\phi\right)^{\frac{(2n+1)\lambda - n}{2n}}}{\left(\int_0^\pi (\sin\phi)^{\frac{2n+1}{n}} d\phi\right)^{\frac{(2n+1)\lambda}{2n}}}\right]\left\{\frac{((s-1)n+1)\lambda + n}{n}\right\}^{\frac{1}{2}} \quad \text{sphere} \tag{41}$$

4. **Result and Discussion**

Table 1 gives the heat transfer rate for different values of $n$ and the dimensionless MHD parameter $N$ for the vertical flat plate at $x = L_r$. The results are presented for isothermal $(\lambda = 0)$ and non-isothermal $(\lambda = 1)$ plate with constant viscosity $(\gamma = 0)$. For the pseudoplastic fluid it is observed that heat transfer rate in the absence of magnetic effect $(N = 0)$ is less than the heat transfer rate in the presence of magnetic effect $(N \neq 0)$. Whereas in the case of Newtonian and dilatant fluid the heat transfer rate decreases as the value of $N$ increase from 0 to 10. This can be attributed to the fact that pseudoplastic fluids are shear-thinning fluids that have a lower apparent viscosity at higher shear rates.

Table 2 gives the heat transfer rate for different values of $n$ and the dimensionless MHD parameter $N$ for the vertical flat plate at $x = L_r$ for $\gamma = 0.5$. The results are presented for isothermal $(\lambda = 0)$ and non-isothermal $(\lambda = 1)$. It is observed that the heat transfer rate decreases as the value of $N$ increases from 0 to 10 for all the three cases, namely, preudoplastic, Newtonian and dilatant fluid for varying viscosity.

Figures 2 – 4 illustrate the effect of non-dimensional MHD parameter $N$ on the heat transfer rate for an isothermal cylinder and sphere. The qualitative behavior of the heat transfer rate is same as that obtained by Bagai and Nishad [19] in the absence of the magnetic field. As the value of $N$ increases the heat transfer rate decreases. Moreover, for the problem under consideration, the fluid heats the isothermal body immersed in it. Figures 5 – 7 depict the effect of the parameter $N$ for a non-isothermal body. It is observed that for higher values of $N$ the body is still heated by the fluid whereas for lower values of $N$ the body heats the fluid. There will be a critical value of $N$, between 2.0 and 5.0, for which there will be negligible heat transfer even though the temperature of the body is varying. For a non-isothermal body $(\lambda = 1)$, it is observed that the local heat flux has a point of extrema between the lower and the upper stagnation points. But as the value of $n$ increases the point of maxima shifts towards the lower stagnation point.

**Table 1** values of $-\theta'(0)$ for different values of MHD parameter $N$ and fluid index $n$ for isothermal body $(\lambda = 0)$ and non-isothermal body $(\lambda = 1.0)$ when viscosity parameter $\gamma = 0$.

| N | Pseudoplastic fluid $(n = 0.5)$ | | Newtonian fluid $(n = 1.0)$ | | Dilatant fluid $(n = 2.0)$ | |
|---|---|---|---|---|---|---|
| | Isothermal body $(\lambda = 0)$ | Non-isothermal body $(\lambda = 1.0)$ | Isothermal body $(\lambda = 0)$ | Non-isothermal body $(\lambda = 1.0)$ | Isothermal body $(\lambda = 0)$ | Non-isothermal body $(\lambda = 1.0)$ |
| 0 | - 0.2576 | 0.1777 | - 0.2153 | 0.2621 | - 0.1780 | 0.3276 |
| 2 | - 0.1849 | 0.2808 | - 0.4879 | - 0.0006 | - 0.4398 | 0.0594 |
| 5 | - 0.4951 | - 0.0159 | - 0.6114 | - 0.1162 | - 0.5884 | - 0.0859 |
| 10 | - 0.6400 | - 0.1500 | - 0.6949 | - 0.1936 | - 0.6845 | - 0.1776 |

**Table 2** values of $-\theta'(0)$ for different values of MHD parameter $N$ and fluid index $n$ for isothermal body $(\lambda = 0)$ and non-isothermal body $(\lambda = 1.0)$ when viscosity parameter $\gamma = 0.5$.

| N | Pseudoplastic fluid $(n = 0.5)$ | | Newtonian fluid $(n = 1.0)$ | | Dilatant fluid $(n = 2.0)$ | |
|---|---|---|---|---|---|---|
| | Isothermal body $(\lambda = 0)$ | Non-isothermal body $(\lambda = 1.0)$ | Isothermal body $(\lambda = 0)$ | Non-isothermal body $(\lambda = 1.0)$ | Isothermal body $(\lambda = 0)$ | Non-isothermal body $(\lambda = 1.0)$ |
| 0 | 0.0070 | 0.4193 | - 0.0792 | 0.3873 | - 0.1086 | 0.3921 |
| 2 | - 0.2608 | 0.2108 | - 0.4603 | 0.0231 | - 0.4270 | 0.0699 |
| 5 | - 0.5283 | - 0.0455 | - 0.6006 | - 0.1071 | - 0.5859 | - 0.0833 |
| 10 | - 0.6576 | - 0.1655 | - 0.6902 | - 0.1896 | - 0.6839 | - 0.1771 |

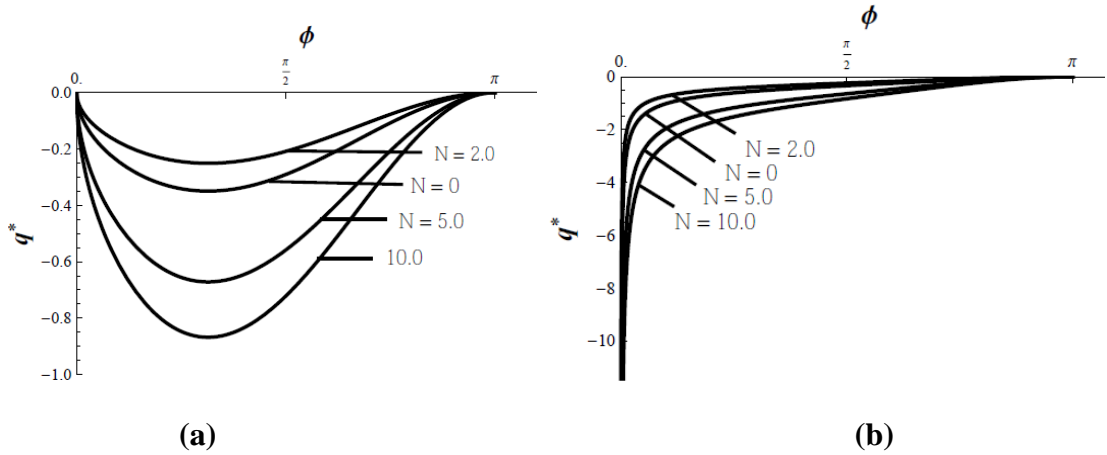

**Figure 2** Effect of MHD parameter $N$ on local heat flux for Pseudoplastic fluid $(n=0.5)$ where $\lambda=0$, $\gamma=0$ for (a) cylinder and (b) sphere.

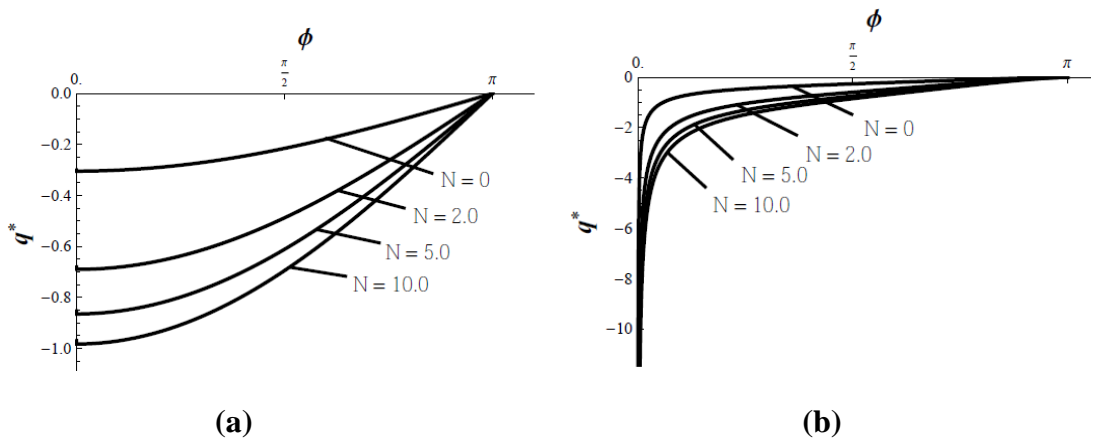

**Figure 3** Effect of MHD parameter $N$ on local heat flux for Newtonian fluid $(n=1.0)$ where $\lambda=0$, $\gamma=0$ for (a) cylinder and (b) sphere.

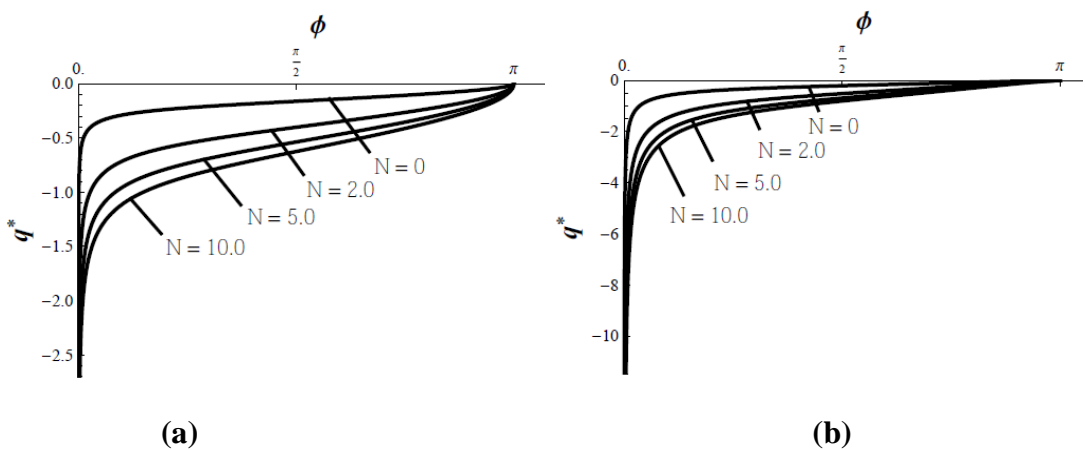

**Figure 4** Effect of MHD parameter $N$ on local heat flux for dilatant fluid $(n=2.0)$ where $\lambda=0$, $\gamma=0$ for (a) cylinder and (b) sphere.

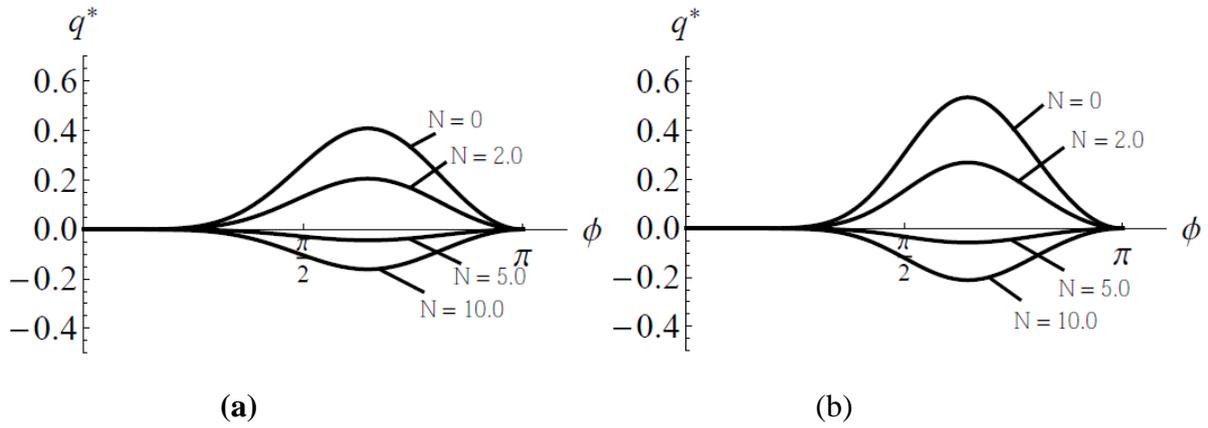

**Figure 5** Effect of MHD parameter $N$ on local heat flux for Pseudoplastic fluid $(n = 0.5)$ where $\lambda = 1.0$, $\gamma = 0.5$ for (a) cylinder and (b) sphere.

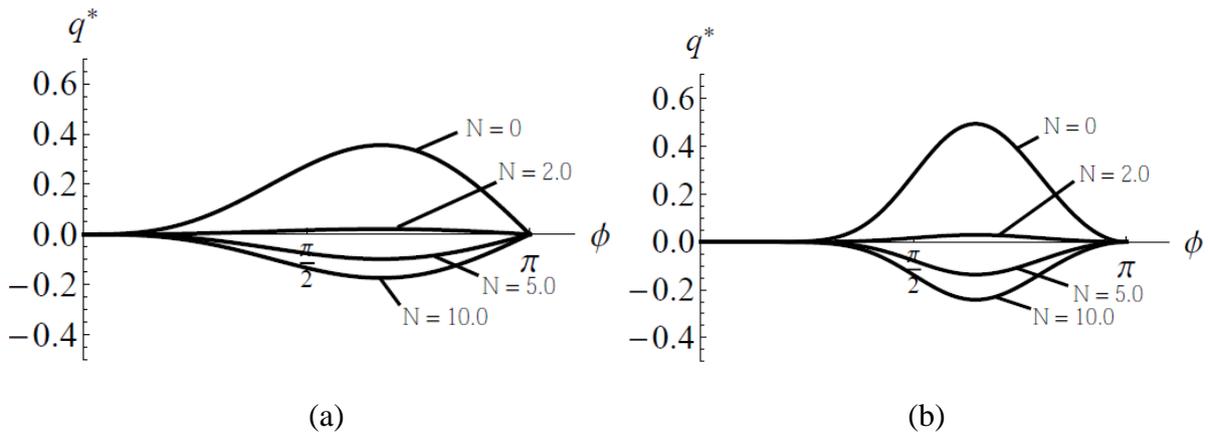

**Figure 6** Effect of MHD parameter $N$ on local heat flux for Newtonian fluid $(n = 1.0)$ where $\lambda = 1.0$, $\gamma = 0.5$ for (a) cylinder and (b) sphere.

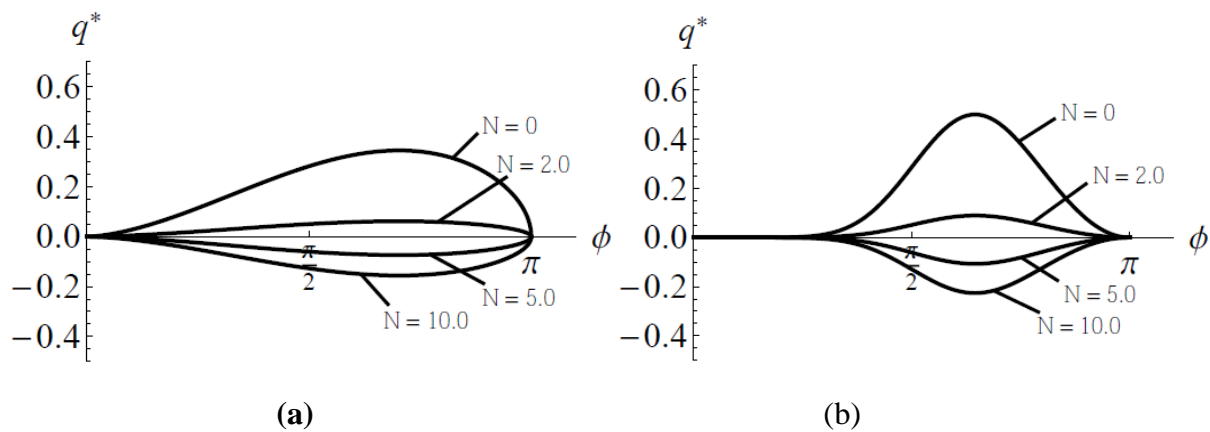

**Figure 7** Effect of MHD parameter $N$ on local heat flux for dilatant fluid $(n = 2.0)$ where $\lambda = 1.0$, $\gamma = 0.5$ for (a) cylinder and (b) sphere.

**List of symbols**

$A(x)$   Function adjusting the general transformation $\left[\alpha_m r^* \left(Ra_x I(x)\right)^{\frac{1}{2}}\right]$

$B(x)$   Function adjusting the general transformation $\left[\dfrac{1}{x}\left(\dfrac{Ra_x}{I(x)}\right)^{\frac{1}{2}}\right]$

$B_0$   Magnetic field strength

$c_{pf}$   Specific heat at constant pressure of the fluid

$f$   Dimensionless stream function

$g, g_x$   Acceleration due to gravity and its stream wise component

$I(x)$   Function adjusting the boundary layer length scale

$k_m$   Effective thermal conductivity of porous medium

$K^*(n)$   Modified permeability

$L_r$   Reference length

$N$   Dimensionless MHD parameter $\left[Ra_x\left(\dfrac{\alpha_m \sigma_0 B_0^2}{x\rho\varepsilon g_x \beta \Delta T_w(x)}\right)\right]$

$n$   Power law index for the fluid

$Nu_x$   Local Nusselt number

$P(x)$   Function associated with wall temperature

$p$   Pressure

$q'''$   Internal heat generation per unit volume

$q_w$   Local surface heat flux

$q^*$   Dimensionless local heat flux

$Ra_x$     Modified local Rayleigh number $\left[\left(\dfrac{K^*(n)g_x\beta\Delta T_w(x)x^n}{\alpha_m^n \upsilon_0^*}\right)^{\frac{1}{n}}\right]$

$r^*(x), r(x)$ Geometric configuration of the body

$T$     Temperature

$u, v$     Velocity component in $x$ and $y$-direction

$x, y$     Boundary layer co-ordinates

**Greek symbols**

$\alpha_m$    Thermal diffusivity of the porous medium

$\beta$    Coefficient of thermal expansion

$\eta$    Similarity variable

$\theta$    Dimensionless temperature

$\lambda$    Exponent associated with wall temperature increase

$\rho$    Fluid density

$\sigma_0$    Electric conductivity

$\upsilon_0^*$    Kinematic viscosity

$\mu^*$    Dynamic viscosity parameter

$\psi$    Stream function

$\gamma$    Viscosity parameter

$\delta$    Boundary-layer thickness

$\varepsilon$    Porosity

$\xi$    Transformed variable

$\phi$    Peripheral angle measured from the lower stagnation point

**Subscripts**

*w*  Wall condition

∞  Ambient condition

0  The leading edge condition